\documentclass[aps,prd,twocolumn,groupedaddress,showpacs]{revtex4-1}

\usepackage{amssymb}
\usepackage{amsmath}
\usepackage{graphicx}
\usepackage{enumerate}
\usepackage{mathtools}
\usepackage{color}
\usepackage{bbold}
\usepackage{subfigure}
\usepackage{slashed}
\usepackage{xcolor}

\begin{document}


\title{Self-interacting sterile neutrino dark matter: the heavy-mediator case}


\author{Lucas Johns}
\email[]{ljohns@ucsd.edu}
\author{George M. Fuller}
\affiliation{Department of Physics, University of California, San Diego, La Jolla, California 92093, USA}


\date{\today}

\begin{abstract}
For active--sterile mixing to be responsible for the full relic abundance of dark matter additional new physics is needed beyond the keV-scale sterile neutrino itself. The extra ingredient we consider here is the presence of self-interactions among the sterile neutrinos. We examine whether active-to-sterile conversion is amplified enough in this scenario that the observed abundance of dark matter can be obtained with a subconstraint mixing angle. This turns out never to be the case in the region we explore: either self-interactions have too small an impact and cannot escape bounds on the mass and mixing angle, or they have too great an impact and cause dark matter to be overproduced. The sharp transition from marginal to excessive effectiveness occurs because a resonance criterion is met in the effective in-medium mixing angle. Once the system goes resonant the game is as good as over, as nonlinearity in the Boltzmann equation leads to runaway production of sterile neutrinos, beginning at a plasma temperature of a few hundred MeV and typically ending at a few tens of MeV. The scenario is therefore ruled out largely by its own dynamics. In this study we focus exclusively on mediators heavier than $\sim 1$ GeV; future work will extend the analysis to lighter mediators, allowing for contact to be made with the kinds of scenarios motivated by issues of small-scale structure.
\end{abstract}


\maketitle


\section{Introduction}

Dark matter has thus far refused to cooperate with the intense experimental efforts to detect it, inspiring many physicists to broaden the search. A great deal of energy in recent years---sometimes motivated by dark matter, sometimes not---has gone toward sterile neutrinos, toward self-interacting dark sectors, and occasionally toward the overlap. We work here in the spirit of an expansive assessment of dark matter candidates, asking how production is affected when the paradigms of sterile neutrinos and self-interactions intersect.

There is good reason to think nonminimally about sterile neutrino dark matter. The simplest scenario \cite{dodelson1994}, in which sterile neutrinos are the only beyond-Standard-Model (BSM) physics and are populated by their mixing with the active states, is strongly disfavored by observations \cite{kusenko2009, adhikari2017, abazajian2017, boyarsky2019}. X-ray and $\gamma$-ray experiments, which look for monochromatic photons from sterile-neutrino decay, bound the mixing angle from above \cite{abazajian2001, yuksel2008, boyarsky2008, watson2012, ng2015, perez2017, dessert2018, ng2019}, and a number of structure-related probes, including phase space, subhalo counts, the Lyman-$\alpha$ forest, and reionization history, bound the mass from below \cite{boyarsky2009, boyarsky2009b, polisensky2011, horiuchi2014, horiuchi2016, schneider2016, rudakovskyi2016, cherry2017, baur2017, schneider2018, vegetti2018}. The combination of these constraints has made it necessary to look beyond the classic Dodelson--Widrow mechanism if one wants sterile neutrinos to comprise all of the dark matter observed in the universe.

Several alternative ways of producing the sterile neutrinos have been proposed, all sharing in common the fact that they invoke at least one additional piece of new physics aside from the sterile neutrino itself. Some of these scenarios amplify production by methods unrelated to active--sterile mixing, as when sterile neutrinos appear as decay products of another new particle \cite{shaposhnikov2006, petraki2006, merle2014, shuve2014}, when an overabundant population of sterile neutrinos is diluted by new sources of entropy \cite{patwardhan2015}, or when sterile neutrinos undergo thermalization \cite{hansen2017} or a SIMP-like freeze-out \cite{herms2018}. Other scenarios alter the mixing itself, as when the vacuum parameters are mediated by an axionlike \cite{berlin2017} or scalar \cite{bezrukov2017, bezrukov2018} field or when a large cosmic lepton number alters the effective in-medium mixing angle \cite{shi1999, abazajian2001b, kishimoto2006, kishimoto2008, ghiglieri2015, venumadhav2016}. The last of these, known as the Shi--Fuller mechanism, has garnered perhaps the most attention among the alternatives, especially in the years following the first detections of an unidentified X-ray line near 3.5 keV in the spectra of various galaxies and galaxy clusters \cite{bulbul2014, boyarsky2014}. Whether this line can be attributed to the radiative decay of sterile neutrinos remains contentious, but forthcoming instruments with high energy resolution will put it definitively to the test \cite{tashiro2018, barcons2015, neronov2016}. 

We consider an alternative to the Dodelson--Widrow scenario in which production is facilitated by interactions in the sterile sector. The effect of self-interactions is twofold: First, the nonzero interaction rate of sterile neutrinos boosts the rate of decoherence, which in turn enhances the transition rate from active to sterile. Second, the nonzero coherent scattering of sterile neutrinos modifies the dispersion relation of neutrinos in the plasma, increasing the effective in-medium mixing angle. The two factors---larger scattering rate and stronger mixing---work in the same direction, suggesting the possibility that, at the expense of introducing self-interactions, the observed abundance of dark matter might be generated with a much smaller vacuum mixing angle than is needed in Dodelson--Widrow. The similarities with Shi--Fuller, moreover, suggest that self-interacting sterile neutrinos might even be consistent with the 3.5 keV line, exchanging the lepton number for a new coupling. The point of this paper is to evaluate these suspicions.

More generally, self-interactions of dark matter are under intensely active investigation because of their possibly ameliorative influence on small-scale structure \cite{tulin2018, arguelles2016}. While we are interested in making contact with this body of work, we will not be able to do so here because we focus exclusively on the limit in which the new mediator is very heavy, an assumption that simplifies the analysis in a number of ways. Keeping the coupling perturbative, while at the same time staying in the heavy-mediator limit, precludes any consideration of the large cross sections needed to hold sway over the dynamics of dark matter halos. The tantalizing case of lighter mediators is left for future work. We settle here for making some brief remarks in the conclusion on how that analysis is expected to differ from the present one.

It is also worth noting that self-interactions among sterile neutrinos have been discussed in connection to the persistent anomalies in short-baseline oscillation experiments \cite{dasgupta2014, hannestad2014, bringmann2014, ko2014, chu2015, archidiacono2015, mirizzi2015b, tang2015, archidiacono2016, archidiacono2016b, cherry2016, forastieri2017, bakhti2017, song2018, jeong2018, chu2018}. These are sterile neutrinos of a different variety, being at the eV scale and therefore much too light to be of relevance to dark matter. In fact, the problem facing eV sterile neutrinos is somewhat like the reverse of the problem facing those at the keV scale: because experimental fits indicate a small mass and a large mixing angle, the challenge is to \textit{prevent} eV sterile neutrinos from being populated in the early universe. This, indeed, is the purpose for which self-interactions are invoked. But despite the difference in model-building philosophy, the underlying physics is closely related.

One last tie-in deserves mention. If they exist, sterile neutrinos at the MeV scale and below are not only frozen into the early universe but are also, much later, produced and emitted by core-collapse supernovae. This includes, of course, the keV dark matter contenders, whose creation benefits from the active neutrinos encountering at least one resonance on their way out of the proto-neutron star, as in Refs.~\cite{abazajian2001b, hidaka2006, hidaka2007, warren2016}. Formulating accurate constraints on the basis of sterile neutrino production in supernovae is a challenge, made even more so if the particles are self-interacting. We do not take up the task here, but we refer to Ref.~\cite{arguelles2019} for a recent analysis of the standard scenario, where sterile neutrinos are truly inert except for their mixing.

\begin{figure}
\centering
\includegraphics[width=.45\textwidth]{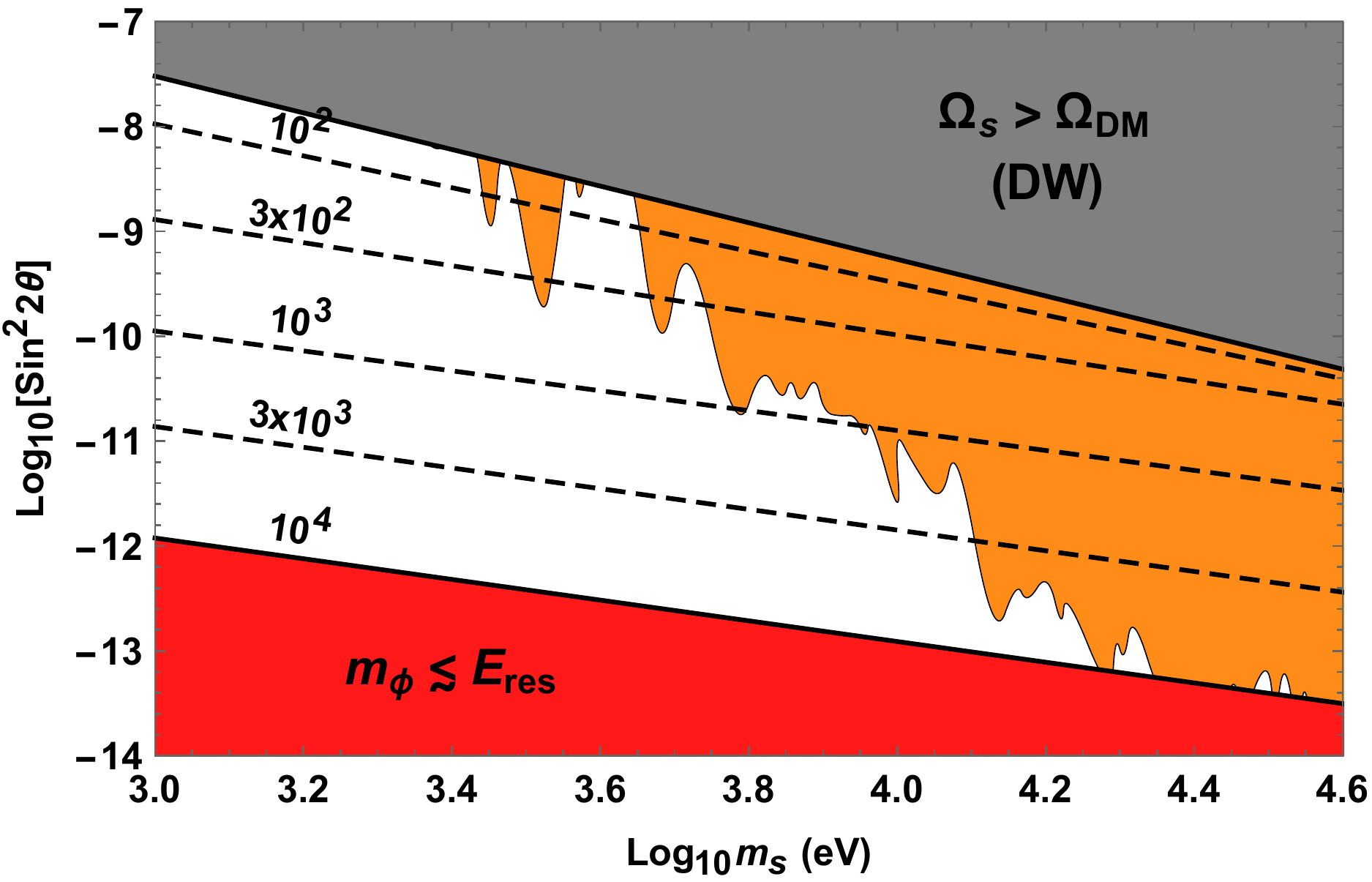}
\caption{Curves indicate the vacuum mixing angles above which sterile neutrinos are overproduced ($\Omega_s > \Omega_\textrm{DM}$) for various choices of self-interaction strength $G_\phi$ (up to $10^4 G_F$, past which either the heavy-$\phi$ assumption breaks down or the coupling becomes nonperturbative). The Dodelson--Widrow mechanism produces $\Omega_s = \Omega_\textrm{DM}$ along the solid line bordering the gray region. X-ray and $\gamma$-ray constraints (orange) \cite{boyarsky2008, horiuchi2014, perez2017, ng2019} are plotted to orient the overproduction curves relative to bounds from radiative decay.}  
\label{constraints}
\end{figure}

\begin{figure*}
\centering
\begin{subfigure}{
\centering
\includegraphics[width=.45\textwidth]{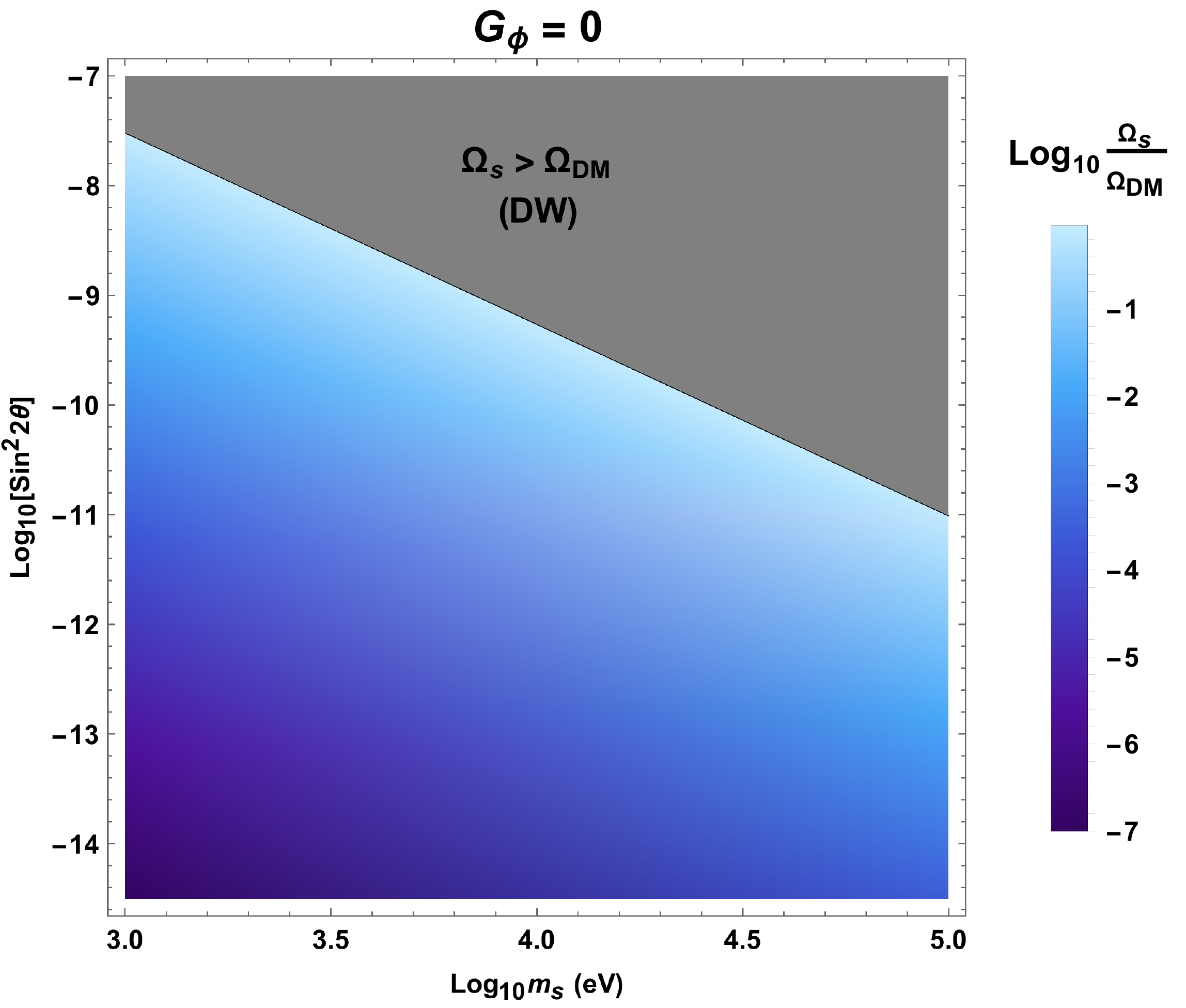}
}
\end{subfigure}
\begin{subfigure}{
\centering
\includegraphics[width=.45\textwidth]{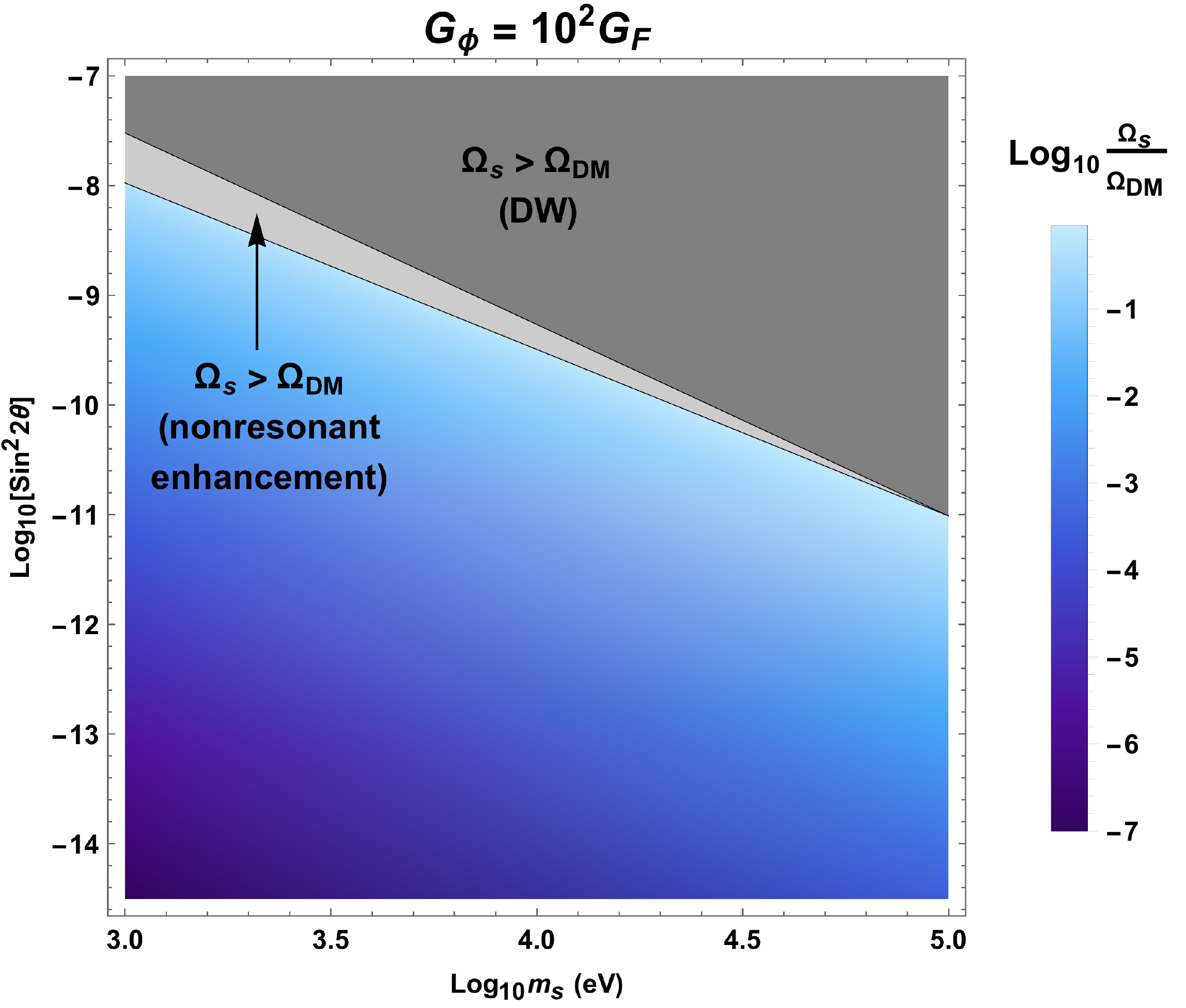}
}
\end{subfigure}

\begin{subfigure}{
\centering
\includegraphics[width=.45\textwidth]{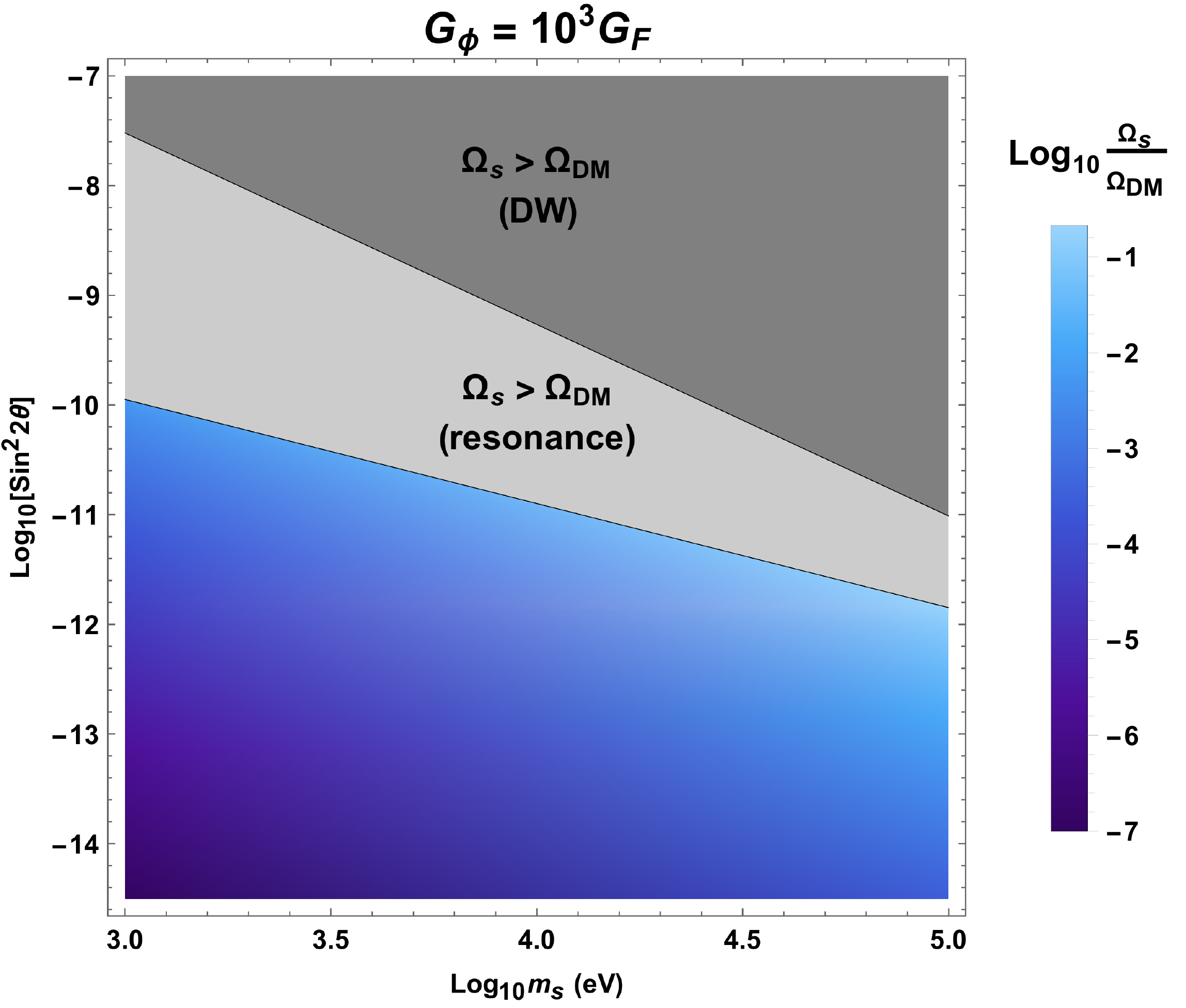}
}
\end{subfigure}
\begin{subfigure}{
\centering
\includegraphics[width=.45\textwidth]{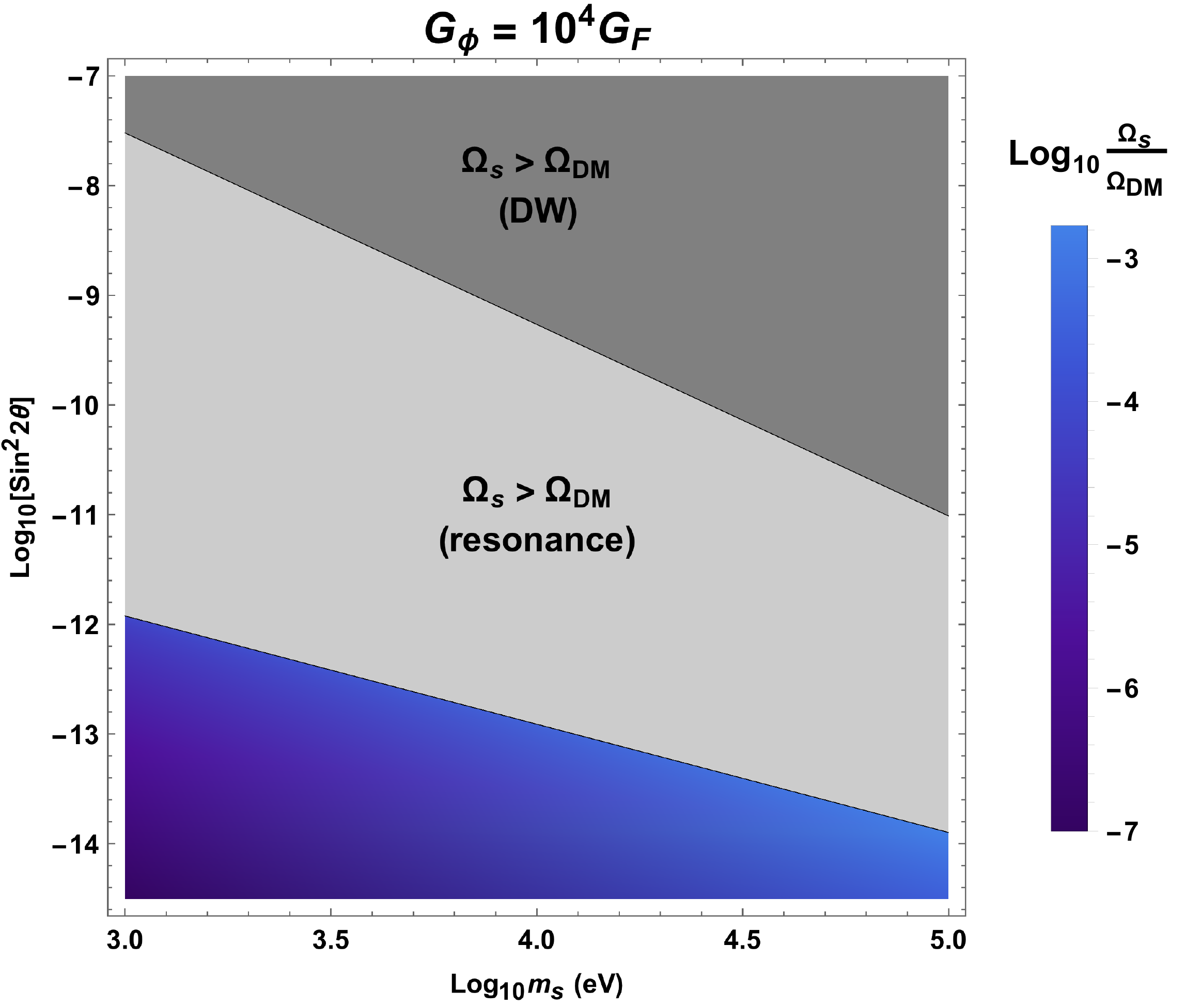}
}
\end{subfigure}
\caption{Fraction of relic sterile neutrino density $\Omega_s$ to observed dark matter density $\Omega_\textrm{DM}$. Dark gray indicates overproduction due to the Dodelson--Widrow mechanism, light gray indicates overproduction due to self-interactions. For $G_\phi = 10^2 G_F$, $\Omega_s / \Omega_\textrm{DM} = 1$ is achieved at slightly smaller mixing compared to $G_\phi = 0$ because $\Gamma_\textrm{tot}$ is slightly larger and $\theta_m$ is nonresonantly enhanced. For $G_\phi = 10^3 G_F$ and $G_\phi = 10^4 G_F$, $\Omega_s / \Omega_\textrm{DM}$ never reaches unity because resonant enhancement sets in.  }
\label{colors}
\end{figure*}

In the rest of this paper we study whether self-interactions are a viable way to rescue sterile neutrino dark matter from current bounds on the mass and mixing. In the heavy-mediator limit, the answer is a flat no, as the factors poised to abet production ultimately conspire to make self-interactions far too much of a good thing. The central finding is that, for any choice of coupling, the $(m_s, \theta)$ parameter space is split into two regions, one where the effect of self-interactions is only marginal and one where it is overwhelming (Figs.~\ref{constraints} and \ref{colors}). The difference between these regions is whether the active--sterile mixing ever becomes resonant. As we show below, both numerically and analytically, resonance is guaranteed whenever the rate of self-interactions is large enough to be very impactful---and, because the Boltzmann equation is nonlinear in the density of sterile neutrinos, runaway production inevitably results. Even fine-tuning the parameters is to no avail, since the transition between these regions is a sharp one. In the end, either dark matter is severely underproduced or it is severely overproduced.

In the next section we set up the equations governing sterile-neutrino production, discuss the underlying physics, and introduce the model used in the calculations. In Sec.~\ref{relic} we present the results, showing that self-interacting sterile neutrinos cannot be all of the dark matter if their interactions are mediated by a very heavy particle. In Sec.~\ref{conclusion} we conclude and reflect on how the analysis changes if the mediator is made lighter. The Appendix contains a few notes on the calculation of the sterile-neutrino scattering rate.

\section{Production mechanism and particle model}

If the Standard Model (SM) neutrinos mix with a sterile state, then the propagating modes in the cosmic plasma are active--sterile mixtures, with lifetimes that are finite due to interactions in the medium. Decay of these quasiparticles---or, in other words, flavor decoherence of the propagating modes---is what sources the sterile neutrinos that accumulate in the early universe. In the Dodelson--Widrow scenario, only SM couplings contribute to the in-medium active--sterile mixing and decoherence rate. In a scenario with self-interacting sterile neutrinos, the new coupling contributes as well. As is typical, we assume that no sterile neutrinos inhabit the universe prior to their creation through this mechanism.

Letting $\Gamma_{\textrm{tot}} = \Gamma_a + \Gamma_s$ be the sum of the interaction rates of active and sterile neutrinos, the Boltzmann equation for the sterile neutrino distribution function $f_s (p,t)$ is then
\begin{align}
\frac{\partial f_s}{\partial t} - Hp \frac{\partial f_s}{\partial p} = \frac{\Gamma_{\textrm{tot}}}{2} \frac{\sin^2 2\theta_m}{1+\left( \frac{\Gamma_{\textrm{tot}}}{2 \omega_m} \right)^2} \left( f_a - f_s \right) + \mathcal{C}_s,
\label{boltzmann}
\end{align}
where all variables tacitly depend on neutrino momentum $p$ and time $t$. The functional $\mathcal{C}_s$, which depends on $f_s$ of all momenta, denotes the collision integrals for all-sterile scattering processes; $H$ is the Hubble parameter; and the subscript $m$ indicates that in-medium values are used for the mixing angle and oscillation frequency. In terms of the vacuum mixing angle $\theta$ and the vacuum oscillation frequency $\omega = \delta m^2 / 2p$, the defining formulae are
\begin{align}
\omega_m^2 = \omega^2 \sin^2 2\theta + \left( \omega \cos 2\theta - \mathcal{V} \right)^2
\end{align}
and
\begin{align}
\omega_m^2 \sin^2 2\theta_m = \omega^2 \sin^2 2\theta.
\end{align}
The potential $\mathcal{V}$, also a function of $p$, is generated by forward scattering of neutrinos on particles in the medium. To be consistent with previous studies \cite{abazajian2001b, kishimoto2008, venumadhav2016}, we take $\nu_a$ to be a muon neutrino. Muons are then the relevant charged-lepton population, with total ($\mu^+$ and $\mu^-$) energy density $\rho_\mu$. The potential $\mathcal{V} = \mathcal{V}_\mu + \mathcal{V}_a + \mathcal{V}_s$ is then composed of
\begin{equation}
\mathcal{V}_\mu = - \frac{8 \sqrt{2} G_F}{3 m_W^2} \rho_\mu p
\end{equation}
from $\nu_a$ scattering on $\mu^\pm$,
\begin{equation}
\mathcal{V}_a = - \frac{8 \sqrt{2} G_F}{3 m_Z^2} \rho_a p \label{va}
\end{equation}
from $\nu_a$ scattering on $\nu_a$, and a contribution $\mathcal{V}_s$ from $\nu_s$ scattering on $\nu_s$. The exact form of this last piece depends on the properties of the mediator of $\nu_s$ scattering. For the model that we study, it is 
\begin{equation}
\mathcal{V}_s = + \frac{G_\phi}{3 m_\phi^2} \rho_s p, \label{spotential}
\end{equation}
valid only when $m_\phi$ is much larger than the typical neutrino energy. The analogue of the Fermi coupling constant is defined as $G_\phi = ( g_\phi / m_\phi )^2$, where $g_\phi$ is the sterile-sector coupling and $m_\phi$ is the mediator mass.

One-loop self-energy diagrams also generate a potential proportional to the difference of the neutrino and antineutrino number densities. Although any asymmetry in the active sector does get partially transferred to the sterile sector, we have confirmed that this potential is always unimportant if the lepton number is comparable to the baryon asymmetry, which we assume to be true. If the lepton number is much larger, then the physics explored here will interact in complicated ways with the Shi--Fuller mechanism and with flavor evolution in the active sector \cite{dolgov2002, abazajian2002, wong2002, mangano2005, pastor2009, johns2016, grohs2017, johns2018}.

The scattering rate of muon neutrinos can be written in the form
\begin{equation}
\Gamma_a = c(p, T) G_F^2 T^4 p, \label{eqgammaa}
\end{equation}
where $c(p,T)$ is a momentum- and temperature-dependent coefficient. In our calculations we use the results of Venumadhav et al. \cite{venumadhav2016}, who computed $c(p, T)$ over the range of temperatures relevant to sterile-neutrino production, accounting for the changing degrees of freedom through the quark-hadron transition. We also employ their tabulated data for the relativistic degrees of freedom $g_*$ and $g_{*S}$, which appear in $H$ and in the relation between time and temperature.

For the calculations that follow, we adopt a simple model in which the sterile neutrino $\psi_s$ couples to a heavy real scalar $\phi$:
\begin{align}
\mathcal{L}_s = & \frac{1}{2} \bar{\psi}_s \left( i \slashed{\partial} - m_s \right) \psi_s + \frac{1}{2}( \partial_\mu \phi )^2 \notag \\
&- \frac{1}{2} m_\phi^2 \phi^2 - \frac{g_\phi}{2} \bar{\psi}_s \psi_s \phi. \label{lagrangian}
\end{align}
As we see in the next section, self-interactions facilitate active--sterile conversion through a series of resonances beginning at a temperature $T_\textrm{res}$. For $\phi$ to qualify as heavy, it must have a mass $m_\phi \gg E_\textrm{res}$, where $E_\textrm{res} \sim 3 T_\textrm{res}$, to ensure that Eq.~\eqref{spotential} is valid at the onset of resonance. In practice this means that $m_\phi$ must be at least $\sim 1$ GeV.

The $\nu_s$ scattering rate $\Gamma_s$ is
\begin{equation}
\Gamma_s \approx 3 \times 10^{-2} \alpha G_\phi^2 T^4 p. \label{eqgammas}
\end{equation}
where $\alpha$ is a normalization constant appearing in the ansatz $f_s (p) \simeq \alpha f_\textrm{FD} (p)$, $f_{\textrm{FD}}$ being the thermal Fermi--Dirac spectrum. Taking $f_s$ to have this form is a reasonable approximation that makes it possible to parametrize $\Gamma_s$ in a form similar to $\Gamma_a$. The other approximations implicit in Eq.~\eqref{eqgammas} are noted in the Appendix. We assume that $\Gamma_s$ never becomes so large that the deviations of $f_a$ from equilibrium are important. This assumption has the potential to break down near resonance, since the factor of $\sin^2 2\theta_m$ in Eq.~\eqref{boltzmann} fails to significantly suppress the $f_a$ depletion rate. We will find in the next section that the approximation is justified nonetheless. The fractional change in $f_a$ that occurs over a weak-interaction time scale is always small in the parameter space that we explore.

In addition to enhancing active-to-sterile conversion, self-interactions also modify the thermodynamics of the sterile population. The 2-to-2 process $\nu_s\nu_s \rightarrow \nu_s\nu_s$ kinetically equilibrates the sterile neutrinos provided that $\Gamma_s \gtrsim H$. If self-interactions occur rapidly enough to have a substantial impact on production (that is, if $\Gamma_s \gtrsim \Gamma_a$), then they are guaranteed to be rapid enough to cause kinetic equilibration, by the fact that $\Gamma_a \gg H$ at the temperatures that concern us.

Less obvious is whether the higher-order 2-to-4 and 4-to-2 processes are fast enough to cause chemical equilibration. Dimensionally, one expects $\Gamma_{2\rightarrow 4} \sim G_\phi^2 T^4 \Gamma_s$. Using the assumption $\Gamma_s \gtrsim \Gamma_a$ and the approximation $H \sim T^2 / m_{Pl}$, the condition $\Gamma_{2\rightarrow 4} \gtrsim H$ at $T \sim 100$ MeV translates to $G_\phi \gtrsim \mathcal{O}( 10^4 ) G_F $, which is the upper limit of what we study in this paper. Since sterile-neutrino equilibration does not feed back into production in any considerable way, and since number-changing processes are only expected to be important in a region of parameter space that we find to be ruled out regardless of their presence, we ignore these effects in the results that follow (\textit{i.e.}, $\mathcal{C}_s = 0$ in Eq.~\eqref{boltzmann}). We have checked our results against those obtained when approximate expressions are used for the rates of number-changing interactions, finding that our conclusions are unaltered. Number densities are enhanced by chemical equilibration at large $G_\phi$, but the only cases in which this effect changes the subsequent course of production are those in which the dark matter abundance is overproduced regardless.

\section{The relic density \label{relic}}

By numerically solving the Boltzmann equation, we find that self-interactions facilitated by a heavy mediator are unable to rescue sterile neutrino dark matter from constraints. Either self-interactions have too small an impact and are unable to move production out of the observationally excluded region, or they have too great an impact and elicit excessive production. The reason is that for any choice of $G_\phi$, $m_\phi$, and $m_s$, there is some critical vacuum mixing angle $\theta_c$ above which a resonance criterion is satisfied. Whether the mixing angle is above or below $\theta_c$ makes a radical difference in the dynamics and outcome of production.

The curves in Fig.~\ref{constraints} represent the mixing angles above which $\Omega_s > \Omega_\textrm{DM}$, for various choices of $G_\phi$ (fixing $g_\phi = 0.5$). The curves move progressively downward until $G_\phi$ tops out at $\sim 10^4 G_F$, past which the heavy-$\phi$ assumption begins to be violated. (Alternatively, $g_\phi$ must become nonperturbative if $\phi$ is to remain heavy beyond $\sim 10^4 G_F$.) The orange region marks the part of parameter space excluded by X-ray and $\gamma$-ray observations assuming that sterile neutrinos are all of the dark matter \cite{boyarsky2008, horiuchi2014, perez2017, ng2019}, and the gray region marks the part excluded by overproduction of sterile neutrinos solely through the Dodelson--Widrow mechanism. To be clear, the points within these regions are not excluded a priori in the self-interacting model; they are only necessarily excluded if $G_\phi$ is chosen such that the produced density of sterile neutrinos matches or exceeds the observed density of dark matter. 

Other constraints could be drawn on the plot, including upper bounds on $m_s$ from Milky Way satellite counts or Lyman-$\alpha$ observations, which in the Dodelson--Widrow scenario severely limit the open window in Fig.~\ref{constraints}  \cite{boyarsky2009, boyarsky2009b, polisensky2011, horiuchi2014, horiuchi2016, schneider2016, rudakovskyi2016, cherry2017, baur2017, schneider2018, vegetti2018}. But the final spectrum---on which these constraints depend---is parameter-dependent in the self-interacting model and generally differs from either a Dodelson--Widrow spectrum or a thermal one. If self-interactions were enabling the production of the observed dark-matter density well below the Dodelson--Widrow curve (solid black in Fig.~\ref{constraints}, bordering the gray region), then a careful analysis of the resulting spectrum and its effects on structure would be warranted. This is especially true since number-changing processes might come into play at stronger couplings, thereby causing sterile neutrinos to proliferate and cool and causing structure-related constraints to weaken. Based on our results, however, such an analysis does not appear to be necessary, and the main role of pre-existing bounds on $m_s$ is only to disfavor the smallest values of $G_\phi$, namely those for which $\theta_c$ lies above the mixing angle required by Dodelson--Widrow. At these couplings ($G_\phi \lesssim 10^2 G_F$), we expect the constraints to apply approximately as they do in the absence of self-interactions.

While Fig.~\ref{constraints} locates the overproduction curves relative to radiative-decay constraints, Fig.~\ref{colors} shows that their deeper significance depends on the self-interaction strength. At large couplings, the curves signal sharp transitions from a production regime in which the sterile-neutrino density $\Omega_s$ is much less than the observed dark-matter density $\Omega_\textrm{DM}$, to one in which it is much greater. This is true of $G_\phi = 10^3 G_F$ and $G_\phi = 10^4 G_F$, for which the fraction $\Omega_s / \Omega_\textrm{DM}$ only reaches about $10^{-1}$ and $10^{-3}$, respectively, before the resonance threshold is crossed. The $G_\phi = 0$ (Dodelson--Widrow) panel, in contrast, depicts the fraction smoothly passing through unity. Only in the vicinity of $G_\phi = 10^2 G_F$ do self-interactions allow for $\Omega_s / \Omega_\textrm{DM} = 1$ to be achieved with a mixing angle smaller than in the Dodelson--Widrow scenario, and even then the effect is likely too small to evade constraints. Although not shown in the figure, $\sin^2 2\theta_c$ in this case nearly coincides with the Dodelson--Widrow curve: low enough to have a visible impact, but high enough not to induce a resonance before all of the observed abundance is made. The message, ultimately, is that there is very little leeway for self-interactions to assist in production without overdoing it.

\begin{figure}
\centering
\includegraphics[width=.45\textwidth]{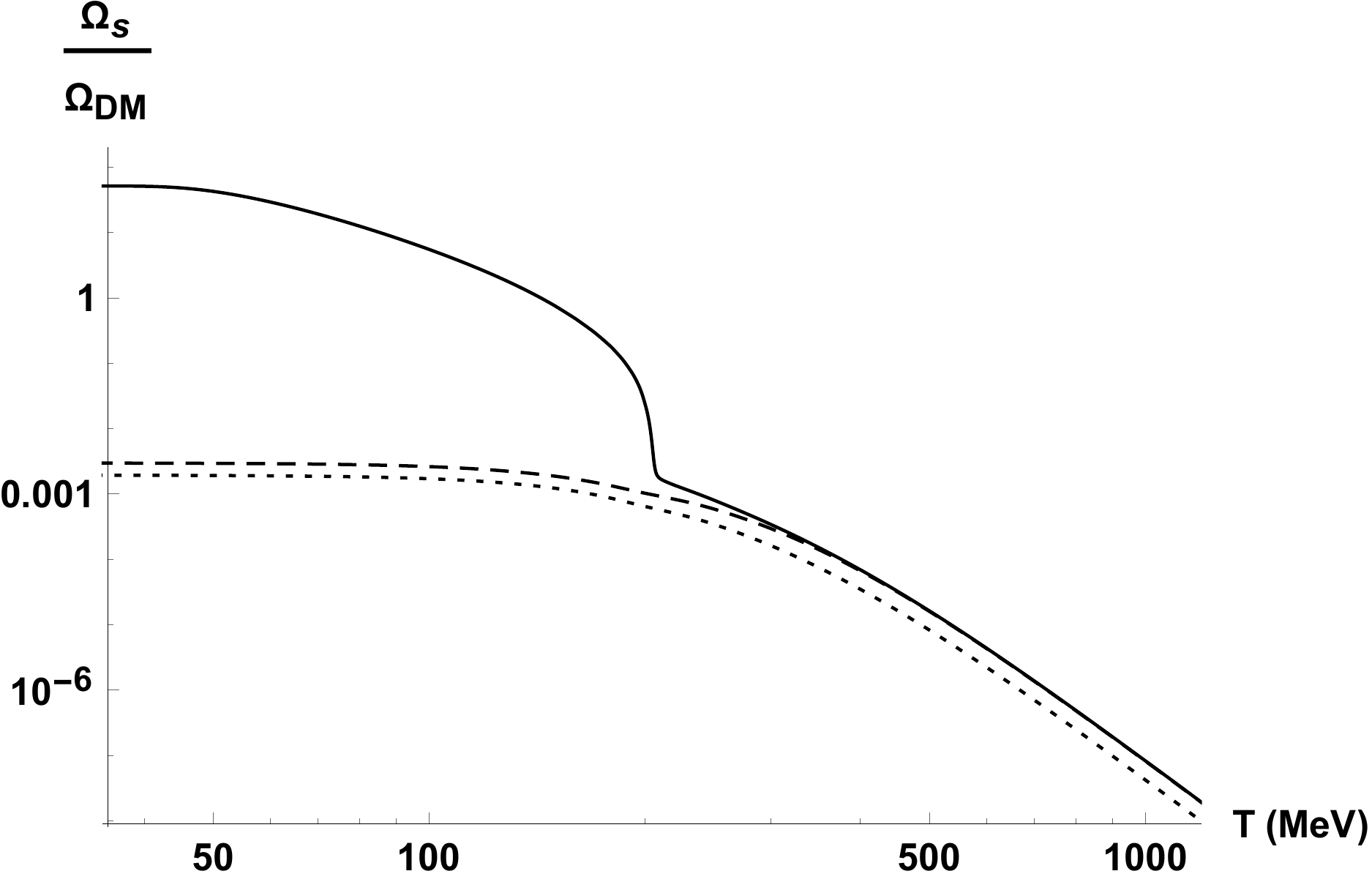}
\caption{$\Omega_s / \Omega_\textrm{DM}$ as a function of temperature. The solid curve corresponds to the test case described in the text (and plotted in Figs.~\ref{growth} through \ref{spectra} as well): $m_s = 10$ keV, $G_\phi = 3 \times 10^3 G_F$, $\sin^2 2 \theta = 2 \times 10^{-12}$. The dashed curve has the same parameters but with $G_\phi = 0$. The dotted curve has the same parameters as the solid curve (including $G_\phi = 3 \times 10^3 G_F$) but with $\sin^2 2 \theta = 1 \times 10^{-12}$, which lies below the critical mixing angle required for resonant production.}  
\label{density}
\end{figure}

\begin{figure}
\centering
\includegraphics[width=.45\textwidth]{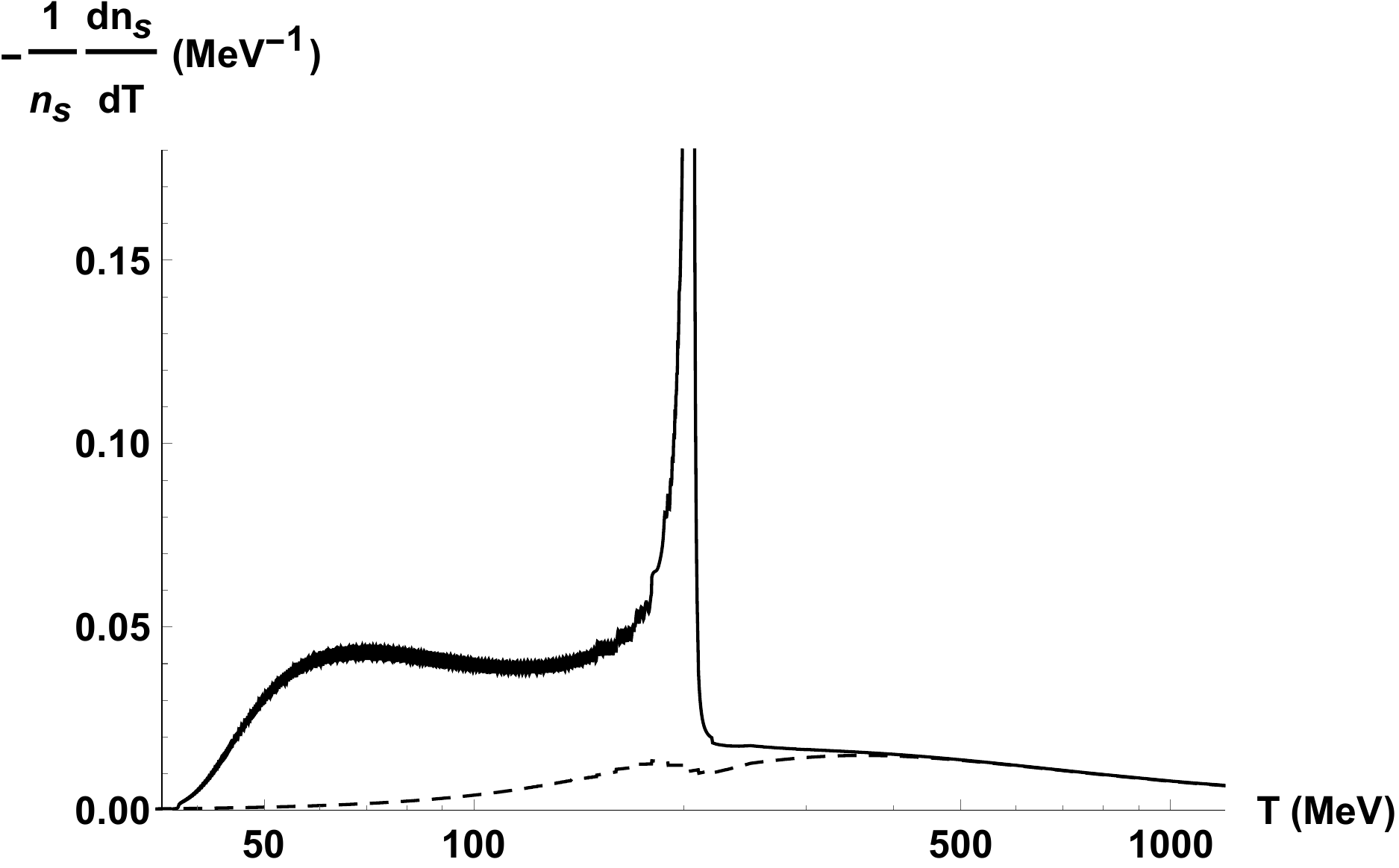}
\caption{Logarithmic growth rate in sterile neutrino number density $n_s$. The dashed curve has the same mixing parameters as the test case (solid curve) but with $G_\phi = 0$. The solid curve peaks at a value of $\sim 0.5$---well off the plot---right when resonant production first sets in, and remains elevated while the resonance sweeps down, and then back up, the neutrino spectrum.}  
\label{growth}
\end{figure}

\begin{figure*}
\centering
\begin{subfigure}{
\centering
\includegraphics[width=.45\textwidth]{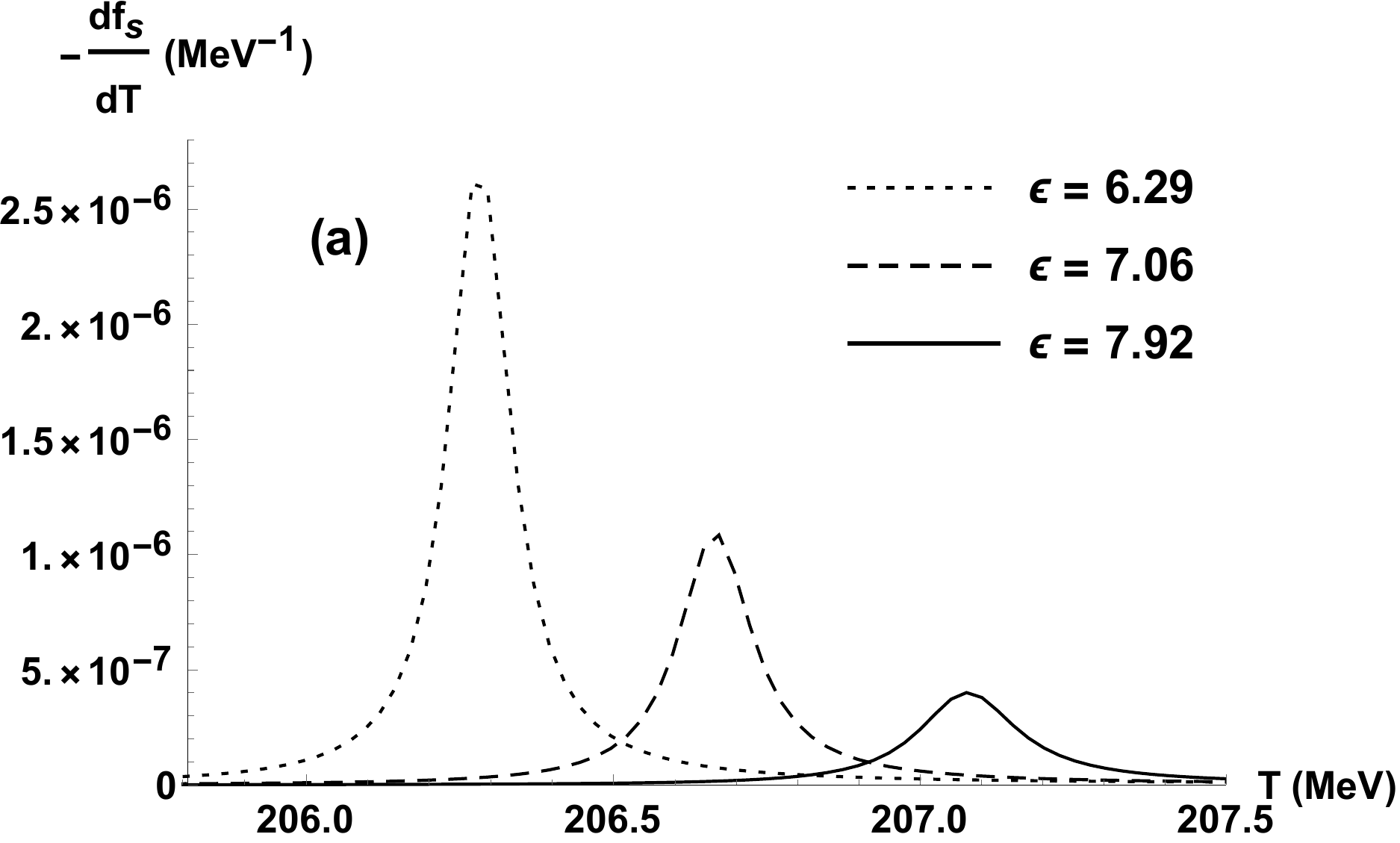}
}
\end{subfigure}
\begin{subfigure}{
\centering
\includegraphics[width=.45\textwidth]{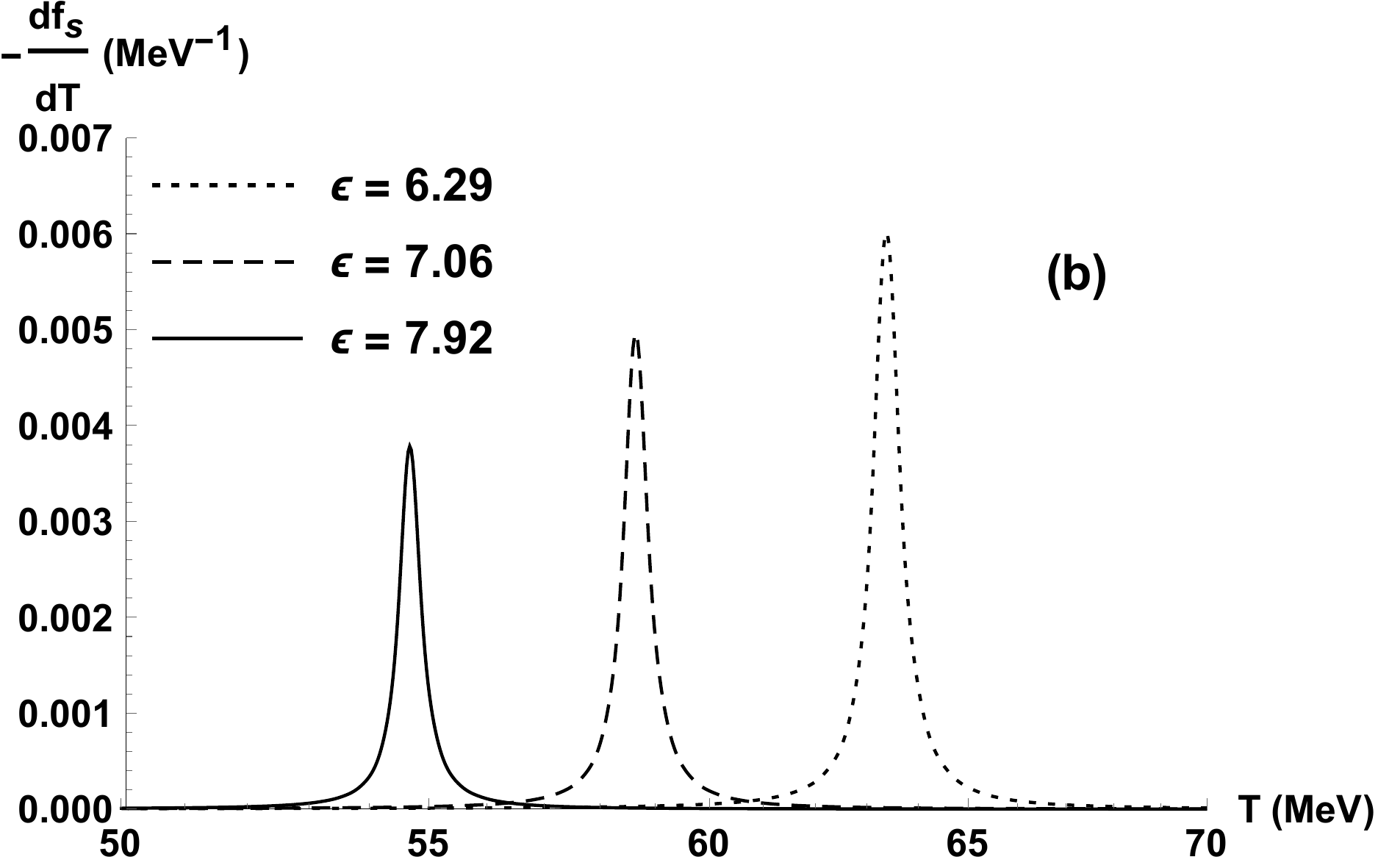}
}
\end{subfigure}
\begin{subfigure}{
\centering
\includegraphics[width=.45\textwidth]{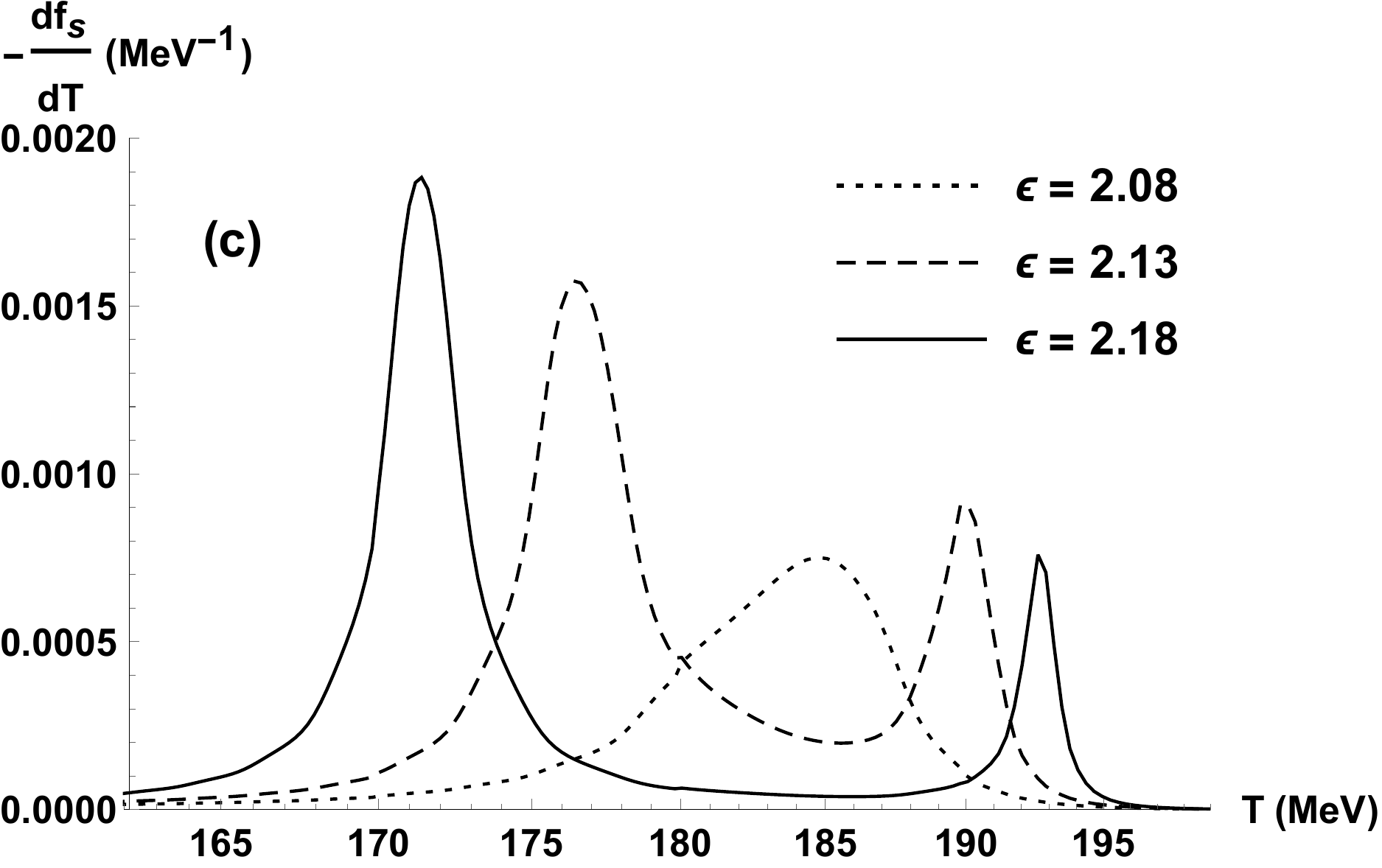}
}
\end{subfigure}
\begin{subfigure}{
\centering
\includegraphics[width=.45\textwidth]{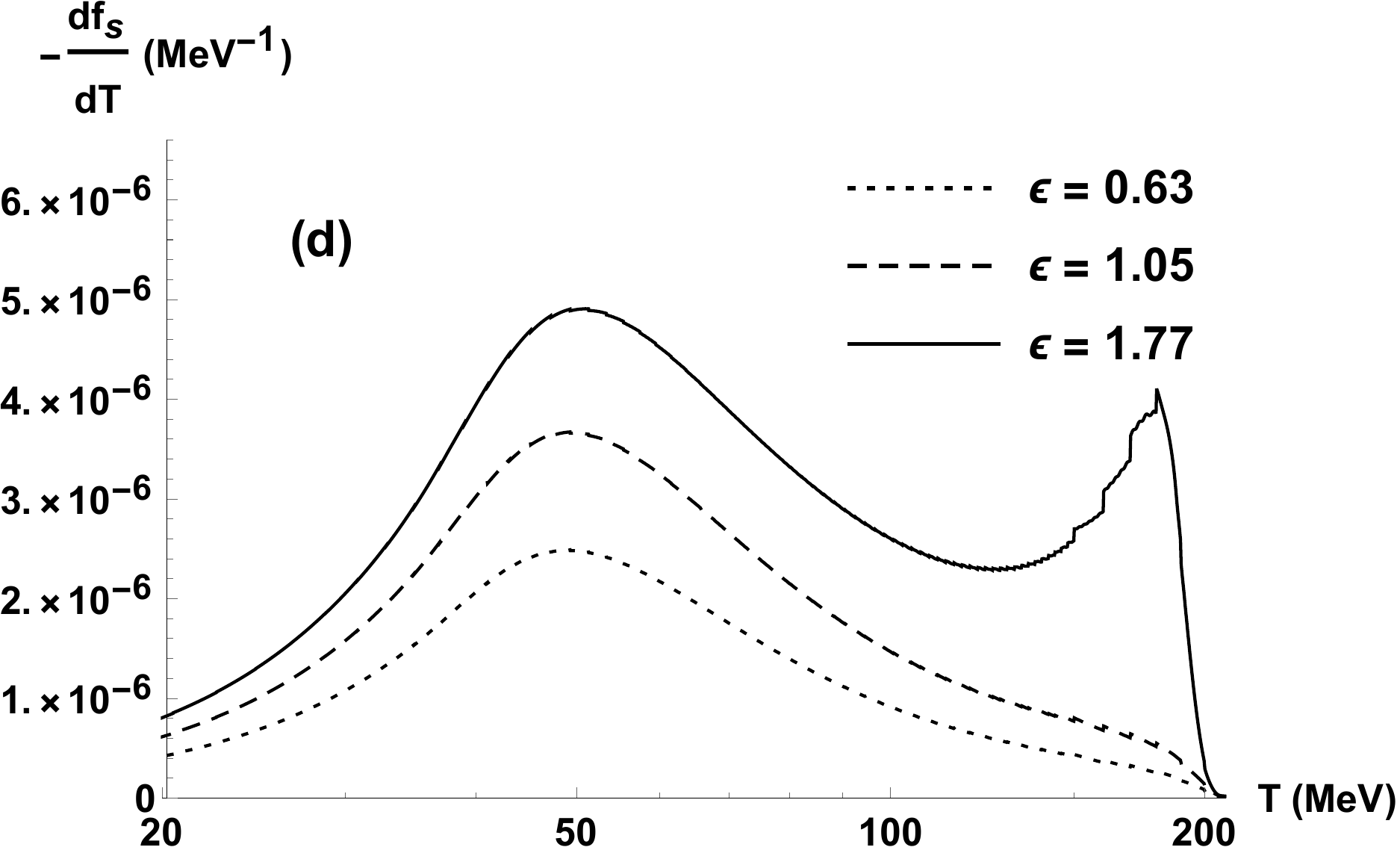}
}
\end{subfigure}
\caption{Growth rate of the sterile neutrino distribution functions due to active--sterile conversion. (a) Neutrinos with energies above the cutoff $\epsilon_\textrm{res}$ go through resonance in quick succession, from high to low energy, leading to a sharp spike in production beginning just above 200 MeV. (b) These neutrinos then pass back through resonance at much lower temperatures, from low to high energy. (c) Neutrinos with energies below $\epsilon_\textrm{burst}$ but above a lower threshold $\epsilon_\textrm{res}$ are pushed through resonance by the burst in production at energies above $\epsilon_\textrm{burst}$. Some of these neutrinos subsequently pass through resonance multiple times; the peaks shown in the panel correspond to these lower-temperature traversals. (d) Neutrinos with energies below $\epsilon_\textrm{res}$ never pass through resonance.}
\label{changes}
\end{figure*}

\begin{figure}
\centering
\includegraphics[width=.45\textwidth]{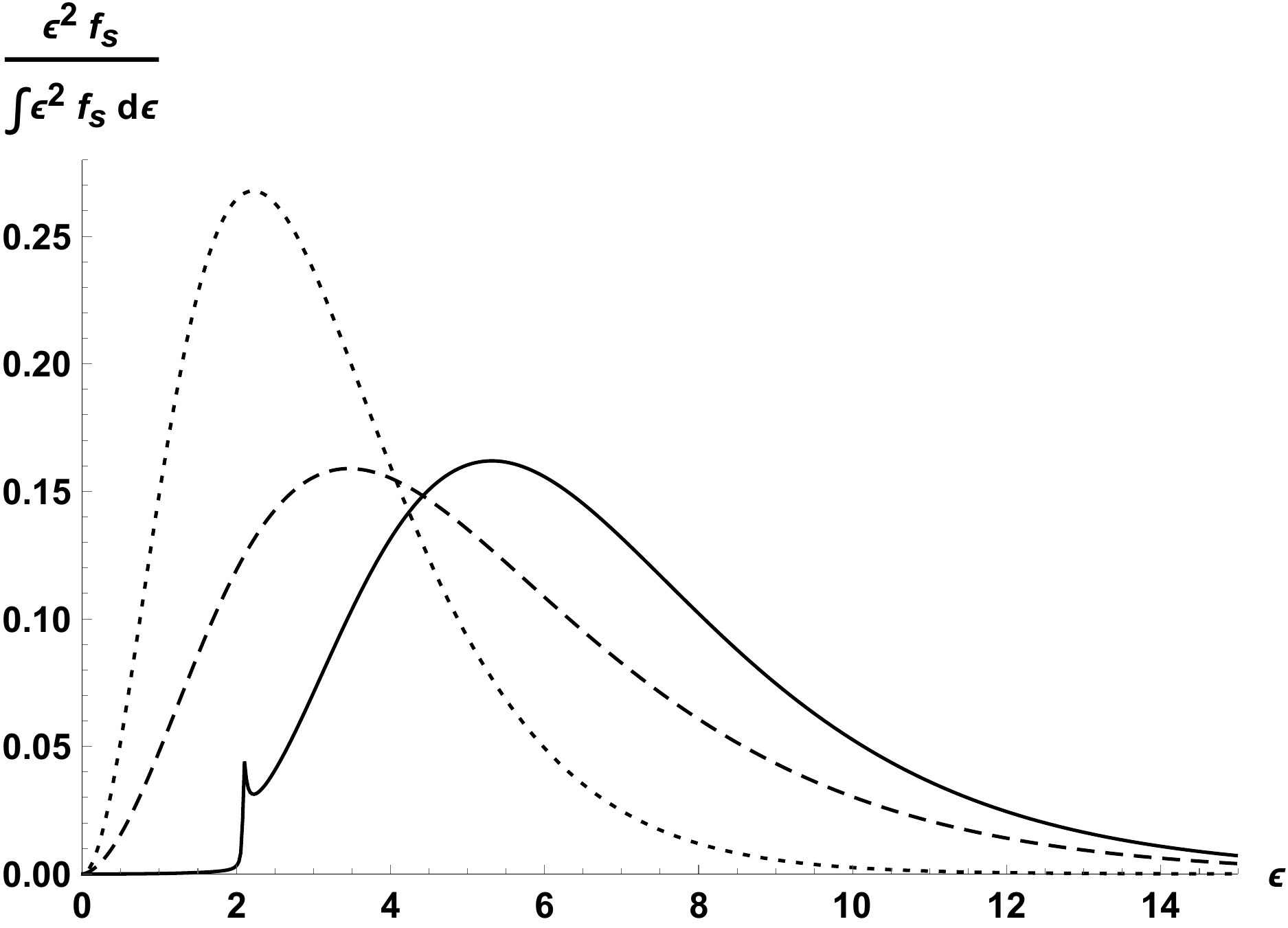}
\caption{Normalized relic spectra, comparing the test case (solid) to Dodelson--Widrow (dashed) and nondegenerate Fermi--Dirac (dotted). The solid curve cuts off sharply at $\epsilon_\textrm{res}$.}  
\label{spectra}
\end{figure}

To zero in on how production changes once $\theta_c$ is surpassed, we shine the spotlight in Figs.~\ref{density} through \ref{spectra} on a single test case: a 10 keV sterile neutrino with $G_\phi = 3 \times 10^3 G_F$ and $\sin ^2 2\theta = 2 \times 10^{-12}$. This mixing angle lies just above $\theta_c$, and as Fig.~\ref{density} shows, the conversion of active neutrinos into sterile ones departs dramatically from what it looks like with the same $\theta$ but $G_\phi = 0$ (dashed curve in the figure) or with the same $G_\phi$ but $\theta < \theta_c$ (dotted curve). At very high temperatures the effect of self-interactions on the abundance is fairly slight, but once the universe cools to $T \sim 200$ MeV, production in the resonant regime (solid curve) suddenly shoots up. After this short-lived period of precipitous fractional growth, $\Omega_s / \Omega_\textrm{DM}$ steadily climbs another four orders of magnitude before being shut off by Hubble expansion. Nonresonant production over the same temperature span is negligible by comparison. Indeed, the similarity in the shapes of the dashed and dotted curves attests to the fact that production in the nonresonant regime is essentially Dodelson--Widrow-like, the normalizations being different only because $\theta$ is.

Like the preceding plot, Fig.~\ref{growth} shows how the sterile-neutrino abundance develops, now presented as a fractional rate of growth. Evident again is the peak right around 200 MeV, which reaches its maximum off the plot. Following the peak, the growth rate oscillates about a track that remains fairly steady, compared to the monotonic decline of the $G_\phi = 0$ curve, down to $\sim 60$ MeV. At $\sim 40$ MeV most of the final abundance has frozen in, but by that point sterile neutrinos already exceed the dark matter density by a factor of $\sim 50$. All of these features---the spike in the growth rate, the subsequent phase during which it remains elevated, and its oscillations---reflect the fact that sterile neutrinos are feeding back into their own production.

With the help of Fig.~\ref{changes}---which shows the growth rates of $f_s (p, T)$ for various neutrino energies and temperature ranges---we can begin to understand the dynamics of the resonant regime. Panel (a) highlights the origin of the peak in Fig.~\ref{growth}: all neutrinos with comoving energies above a cutoff $\epsilon_\textrm{res}$ pass through resonance, one after another, beginning at a temperature just above 200 MeV. Having a sizable fraction of the spectrum go through resonance causes all subsequent evolution of the abundance to differ markedly from Dodelson--Widrow production.

Why resonances are traversed by neutrinos with $\epsilon \gtrsim \epsilon_\textrm{res}$ can be understood as follows. A resonance occurs for neutrinos of a given energy if there is some temperature at which the potential $\mathcal{V}_s$ equals (in magnitude) the SM part of the Hamiltonian. Since $\omega \propto 1 / \epsilon$, this criterion is most easily met by high-energy neutrinos, for which the contribution to the Hamiltonian of $\omega \cos 2\theta$ is small. Taking that term to be negligible, the resonance criterion is $| \mathcal{V}_s | \approx | \mathcal{V}_\mu + \mathcal{V}_a |$, or equivalently
\begin{equation}
\rho_s \approx 8 \sqrt{2} \frac{G_F}{G_\phi} \left( \frac{m_\phi^2}{m_W^2} \rho_\mu  + \frac{m_\phi^2}{m_Z^2} \rho_a  \right). \label{rhoscrit}
\end{equation}
We have seen already that the production of sterile neutrinos prior to the resonance is only marginally enhanced by self-interactions. Put another way, $\rho_s \approx \rho_s^\textrm{DW}$ at these temperatures, where the latter quantity is the energy density of sterile neutrinos generated when $G_\phi$ is set to zero. If this substitution is made on the left-hand side, then Eq.~\eqref{rhoscrit} depends only on self-interaction parameters through their explicit appearance on the right, and the equation becomes a statement about how large $\theta$ must be for the highest-energy neutrinos to have reached resonance at a given temperature. Solving the Boltzmann equation with $G_\phi \neq 0$ is unnecessary for establishing whether a resonance occurs in the system, because the appearance of a resonance depends only (in this approximation) on whether the seed population generated by Dodelson--Widrow production is large enough.

Eq.~\eqref{rhoscrit} is independent of $\epsilon$, meaning that it applies to the neutrino population as a whole. The smallest $\theta$ that satisfies the inequality at any temperature is the critical value $\theta_c$: at and above this mixing angle the system is guaranteed to hit a resonance. The same equation tells us, for $\theta \geq \theta_c$, the temperature $T_\textrm{res}$ at which resonance is first broached. It does not tell us, however, \textit{which} neutrino energies are involved. The value of $\epsilon_\textrm{res}$ cannot be so easily estimated as $\theta_c$ or $T_\textrm{res}$, because as the resonance sweeps downward in energy, $\mathcal{V}_s$ rapidly diverges from the track it follows in the Dodelson--Widrow scenario. In other words, $\epsilon_\textrm{res}$ is set by the nonlinear dynamics of production. We can be sure, however, that resonance will not pass through the \textit{entire} spectrum. Since $\omega \cos 2\theta \rightarrow \infty$ as $\epsilon \rightarrow 0$, there must be some finite cutoff below which $\mathcal{V}_s$ never exceeds (again, in magnitude) the vacuum part of the Hamiltonian.

As the temperature drops, $\omega \cos 2\theta$ begins to dominate over $\mathcal{V}_\mu + \mathcal{V}_a$ even for neutrinos at the high end of the spectrum. But even though $\mathcal{V}_s$ likewise dilutes with five powers of the scale factor, the rapid creation of sterile neutrinos delays the crossing of $\omega \cos 2\theta$ and $\mathcal{V}_s$ till lower temperature. When the crossing does finally occur, neutrinos pass back through resonance, leading to the spikes in production shown in panel (b) of Fig.~\ref{changes}. This time higher-energy neutrinos pass through later than lower-energy ones, as dictated by the scaling of $\omega$ and $\mathcal{V}_s$. In the end, the sweep of resonance across the spectrum is stretched out over a protracted period from $\sim 200$ MeV down to $\sim 40$ MeV. If all energies were instead to pass through resonance in unison, total production would be much more limited. As it is, each resonance takes advantage of the one that preceded it, amplifying the feedback between scattering ($\Gamma_s$) and dispersion ($\theta_m$) and explaining why the magnitudes of production are so much larger in panel (b) than they are in panel (a). Resonant production is self-reinforcing in this way: the growth of $\rho_s$ due to resonant conversion competes against the decline of $\mathcal{V}_s$ due to Hubble expansion, prolonging the sweep from low back up to high energies. And while the growth of $\rho_s$ also \textit{shortens} the initial downward sweep, it compensates by spreading the resonance to more of the spectrum than one would expect without feedback on $\mathcal{V}_s$.

Panel (c) illustrates the dynamics of neutrinos near $\epsilon_\textrm{res} \simeq 2.1$. As $\epsilon$ descends on the cutoff, the two resonance peaks move closer together (solid and dashed curves) until finally merging into one. Just below $\epsilon_\textrm{res}$ (dotted), production remains enhanced by a large $\theta_m$ but never reaches unity. The growth of the resonant peaks as temperature decreases is a reflection of the feedback alluded to in the previous paragraph. Much later the trend reverses, as seen in panel (b), due to the resonance reaching the sparsely populated upper parts of the spectrum.

As shown in panel (d), neutrinos of energies $\epsilon < \epsilon_\textrm{res}$ do not go through resonance at all. Neutrinos in this energy range make a modest contribution to the sterile-neutrino abundance, their production primarily reflecting the scattering rate. The gentle peak near 50 MeV, for example, marks the point at which active--sterile conversion can no longer overcome the redshifting of $\Gamma_s$. Higher-energy neutrinos in this range do see another peak before this one, indicative of the minor enhancement of $\theta_m$ that occurs when sterile neutrinos above $\epsilon_\textrm{res}$ pass through resonance for the first time, but it is pronounced only for energies close to the resonant threshold.

Fig.~\ref{spectra} shows the relic spectrum left over after active--sterile conversion has shut off, juxtaposing the test case (solid) with the Dodelson--Widrow (dashed) and nondegenerate Fermi--Dirac (dotted) spectra. The resonantly produced spectrum is the ``hottest'' of the three, with a negligibly small fraction of number density below $\epsilon_\textrm{res}$. (The small spike right at the cutoff is due to $\epsilon_\textrm{res}$ lingering near resonance while the sweep reverses its direction.) As noted earlier, alterations to the spectrum from sterile-sector scattering---which tend to push it toward an equilibrium distribution---are not included in the calculation.

We have addressed in this section why a series of resonances occurs (in those cases where it does) and what its consequences are. We have not addressed, however, why it is not possible to have $\Gamma_s$ significantly boost production while at the same time avoiding resonance. The explanation is straightforward: whenever $\Gamma_s$ is significant, so is $\mathcal{V}_s$, as shown by the following short argument. If $\Gamma_s \gtrsim \Gamma_a$ at some temperature, then $\alpha \gtrsim ( G_F / G_\phi )^2$ at the same temperature, assuming the numerical coefficients in Eqs.~\eqref{eqgammaa} and \eqref{eqgammas} to be comparable. But since
\begin{equation}
\left| \frac{\mathcal{V}_s}{\mathcal{V}_a} \right| = 8 \sqrt{2} \alpha \frac{G_\phi}{G_F} \left( \frac{m_Z}{m_\phi} \right)^2,
\end{equation}
the lower bound on $\alpha$, along with perturbativity of $g_\phi$, implies that $| \mathcal{V}_s | \gtrsim | \mathcal{V}_a |$.

\section{Discussion \label{conclusion}}

We have studied active--sterile conversion in a model with sterile neutrinos coupled to a new heavy mediator, finding that self-interactions either have very limited impact or cause gross overproduction of dark matter. The essential point, we have seen, is that if self-interactions are strong enough to have a significant effect on the decoherence rate, they are also strong enough to trigger a cascade of resonances in the active--sterile mixing.

For $G_\phi \lesssim \mathcal{O} (10^2 ) G_F$, parameters can be found for which the observed abundance of dark matter is reproduced at a mixing angle smaller than the one required by the Dodelson--Widrow mechanism, but the mixing angle is still not small enough---nor the relic spectrum cold enough---to evade constraints. In the range $\mathcal{O} (10^2 ) G_F \lesssim G_\phi \lesssim \mathcal{O} (10^4) G_F$, resonant production prevents the observed abundance from being reproduced at all. And for $G_\phi \gtrsim \mathcal{O} (10^4) G_F$, the heavy-mediator approximation, which we have used throughout the study, becomes illegitimate. We have observed some hints in our calculations that above $\sim 10^4 G_F$ the correct relic abundance might plausibly be generated for the right choices of parameters. Resonant production still amplifies the active--sterile conversion, just not to excess. But given that the heavy-$\phi$ assumption is dubious in these cases, and given that we are not tracking the effects of number-changing processes, we have chosen, conservatively, to let the $10^4 G_F$ cutoff be a strict one.

Our focus, for the sake of simplicity, has been on a scalar mediator, but the results are expected to be similar for other spins so long as the mass is heavier than the energy scale at which production first becomes appreciable. The differences will be numerical, not qualitative, ones, stemming from the different coefficients in $\mathcal{V}_s$ and $\Gamma_s$. (Compare, for instance, to the formulae in the supplemental materials to Ref.~\cite{dasgupta2014}.)

We have also assumed throughout this paper that the coupling $g_\phi$ is not much smaller than 1. At fixed four-fermion coupling $G_\phi$, smaller $g_\phi$ means smaller $m_\phi$, which in turn means that the system enters resonance more readily ($\mathcal{V}_s \propto 1/m_\phi^2$). We do not find varying $g_\phi$ independently of $G_\phi$ to be of any help in matching the dark matter abundance inferred in the universe.

In theories with keV sterile neutrinos, the primordial plasma is not their only place of origin: supernovae also create them. It is an intriguing question how the constraints apply if self-interactions are involved. Should a large enough seed population of sterile neutrinos be present, the particles may trap themselves and, as in the early universe, trigger a resonance. (Precedents for some of the relevant dynamics can be found in Refs.~\cite{fuller1988} and \cite{konoplich1988}, in the context of neutrino--Majoron couplings.) This line of inquiry will be made especially salient if regimes other than the one studied here are discovered to give rise to the full relic abundance without defying cosmological bounds.

If the goal is to have self-interacting sterile neutrinos make up all of the dark matter, the most promising simple extension of the model studied here is one with a lighter mediator. Masses below $\sim 1$ GeV are small enough that the sterile-sector scattering rates and the oscillation potential are sensitive to the mediator momentum and the presence of an ambient on-shell population. The effect on the potential may be especially important for lighter masses, since $\mathcal{V}_s$ changes sign in passing from $T \gg m_\phi$ to $T \ll m_\phi$.

Aside from having dynamics potentially quite different from the heavy-mediator scenario, models with lighter mediators are also compelling from the standpoint of small-scale structure, which has motivated much of the work on self-interactions. Because even the largest cross sections attainable in the perturbative heavy-mediator limit are still several orders of magnitude too weak to affect halo structure, we cannot yet comment definitively on whether viable regions of parameter space can be found in which halo observations are relevant. Scaling arguments suggest that models with $m_\phi \sim 10^{-3} m_s$---a condition which establishes a velocity-dependence of the cross section that is consistent with observations from dwarf- up to cluster-sized halos \cite{tulin2018}---may be inefficient at converting active neutrinos into sterile ones due to suppression of $\theta_m$, much like what happens to eV sterile neutrinos in Refs.~\cite{dasgupta2014, hannestad2014} and elsewhere. Of course, self-interactions need not alleviate tension at small scales for them to play a decisive part in generating sterile neutrino dark matter. Indeed, halos can just as well be regarded as offering constraints rather than asking for a cure. More work is needed before a comprehensive assessment can be made of sterile-sector interactions on the neutrino portal.

\appendix*
\section{Calculation of $\Gamma_s$ \label{gammas}}

For the process $\nu_s \nu_s \rightarrow \nu_s \nu_s$, the spin-summed square of the amplitude is
\begin{align}
\sum | \mathcal{M} |_{2\rightarrow 2}^2 = 24 \left(\frac{g_\phi}{m_\phi} \right)^4 \bigg[ (p_1& \cdot p_2)^2 + (p_1 \cdot p_3)^2 \notag \\
&+ (p_1 \cdot p_4)^2 \bigg],
\end{align}
where $p_1$ and $p_2$ label the ingoing momenta, $p_3$ and $p_4$ the outgoing. (Unlike in the main text, here we are using $p$ to denote four-momentum, $E$ to denote energy.) Neglecting Pauli blocking, the 2-to-2 scattering rate for a sterile neutrino of momentum $p_1$ is
\begin{align}
\Gamma_s = \frac{1}{8 E_1} \int d\Pi_2 d\Pi_3 d\Pi_4 (2 \pi )^4 \delta^4 &(p_1 + p_2 - p_3 - p_4 ) \notag \\
&\times \sum | \mathcal{M} |_{2\rightarrow 2}^2 f_2,
\end{align}
where $f_2$ is the distribution function of the sterile neutrino with momentum $p_2$ and $d\Pi_i$ is the Lorentz-invariant phase-space volume $d\vec{p}_i^{~3} / (2 \pi)^3 2 E_i$. Since the $(p_i \cdot p_3)^2$ and $(p_i \cdot p_4)^2$ parts become equal once integrated over, only two phase-space integrals need to be computed. We assume that the distribution function of the scatterer is
\begin{equation}
f_2 = \frac{\alpha}{\exp\left( \frac{E_2}{T} \right) + 1}.
\end{equation}
The first of the phase-space integrals then evaluates to
\begin{align}
L_1 &= \int d\Pi_2 d\Pi_3 d\Pi_4 (2 \pi )^4 \delta^4 (p_1 + p_2 - p_3 - p_4 ) \notag \\
&\hspace{1.5in}\times (p_1 \cdot p_2 )^2 f_2 \notag \\
&= \frac{7 \pi}{2880} \alpha E_1^2 T^4.
\end{align}
The second integral can also be done analytically, but the result is a lengthy expression containing polylogarithms of various orders. We coerce it into a form comparable to $L_1$ by setting $E_1 = \langle E_1 \rangle \approx 3.15 T$ and then factoring out two powers of energy:
\begin{align}
L_2 &= \int d\Pi_2 d\Pi_3 d\Pi_4 (2 \pi )^4 \delta^4 (p_1 + p_2 - p_3 - p_4 ) \notag \\
&\hspace{1.5in}\times (p_1 \cdot p_3 )^2 f_2 \notag \\
&\approx 4 \times 10^{-4} ~\alpha E_1^2 T^4.
\end{align}
Combining these,
\begin{equation}
\Gamma_s \approx 0.03 \alpha G_\phi^2 T^4 E_1,
\end{equation}
where $G_\phi = ( g_\phi / m_\phi )^2$.

\begin{acknowledgments}
We thank Baha Balantekin, Susan Gardner, Evan Grohs, Jung-Tsung Li, Amol Patwardhan, and Sebastien Tawa for helpful conversations. This work was supported by NSF Grant No. PHY-1614864 and by the NSF N3AS Hub Grant No. PHY-1630782 and Heising-Simons Foundation Grant No. 2017-228.
\end{acknowledgments}

\bibliography{all_papers}

\end{document}